\documentclass[a4paper,10pt]{article}

\usepackage{geometry}
\makeatletter\@addtoreset{chapter}{part}\makeatother%
\usepackage[dvipdfmx]{graphicx}
\usepackage[normal,oneline]{caption2}
\usepackage{url}
\usepackage[affil-it]{authblk}
\usepackage{textgreek}
\usepackage{xcolor}
\usepackage{bm}
\usepackage{appendix}

\begin{document}

\providecommand{\keywords}[1]
{
  \small	
  \textbf{\textit{Keywords---}} #1
}

\title{\bf Methodology for Modelling the new COVID-19\\
Pandemic Spread and Implementation \\
to European Countries}

\author[]{S.~Maltezos}

\affil[]{National Technical University of Athens\\
Physics Department}

\date{}
\maketitle

\pagenumbering{arabic}

\newcommand{\beq}{\begin{equation}}
\newcommand{\eeq}{\end{equation}}

\newcommand{\ben}{\begin{eqnarray}}
\newcommand{\een}{\end{eqnarray}}

\begin{abstract}
\emph{Abstract.}
After the outbreak of the disease caused by the new virus COVID-19, the mitigation stage has been reached in most of the countries in the world. During this stage, a more accurate data analysis of the daily reported cases and other parameters became possible for the European countries and has been performed in this work. Based on a proposed parametrization model appropriate for implementation to an epidemic in a large population, we focused on the disease spread and we studied the obtained curves, as well as, we investigated probable correlations between the country's characteristics and the parameters of the parametrization. We have also developed a methodology for coupling our model to the SIR-based models determining the basic and the effective reproductive number referring to the parameter space. The obtained results and conclusions could be useful in the case of a recurrence of this repulsive disease in the future. 

\noindent
\end{abstract} \hspace{10pt}

{\keywords{COVID-19, Epidemic, Semi-Gaussian, SIR}}

\section{Introduction}

The disease of the new virus COVID-19 which has been pandemic in the world for about 90 days, the ``wavefront'' of infection has reached its mitigation stage. Therefore, this is the time to turning our thoughts not only to its subsequent, painful and serious implications of this pandemic \cite{Siettos_1}, \cite{Peng}, \cite{Li}, \cite{Liu} but also, it could be useful to analyse the way of growth of the disease among the countries until the $10^{th}$ of May 2020, as well as, to correlate these with the main parameters that likely played a significant role. 

In particular, we consider extremely useful to study the specific characteristics of each country that played a role, the financial level or even genetic behaviour against to the new corona virus and the associated disease. Some of these characteristics used as mathematical parameters for performing correlation studies. The results of this study could give us information for preparing more effective defensive strategies or practical ``tools'' in a possible future return of the pandemic which constitutes the central goal of the present work.
The outline of the paper is as follows. In Section \ref{s2} we present a theoretical methodology for parametrizing an epidemic, in Section \ref{s3} we explain how to couple the present parametrization model with the basic SIR model, in Section \ref{s4} we give results relevant to the end-to-end epidemics growth and in Section 5 we discuss the conclusions.

\section{Theoretical methodology}
\label{s2}
\subsection{Epidemic model}

Our methodology is based on the parametrization of the growth of the COVId-19 disease that we used in our recent work  \cite{Maltezos} and also in \cite{Vazquez} and \cite{Ziff}, that we call ``semi-gaussian of n-degree''. It was used for fitting the disease's growth at various indicative countries and it belongs to the model category of ``Regression Techniques'' for epidemic surveillance. The basic-single term expression of this parametrization model is

\beq
c(t)=A{{t}^{n}}{{e}^{-t/\tau }}
\label{basic_model}
\eeq
\\
where the function $c(t)$ applied in an epidemic spread represents the rate of the infected individuals as the new daily reported cases (DRC) and coincides with the function $I(t)$ in the SIR model, as we can see in the following. Also $A$ is constant while $n$ and $\tau$ are model parameters.  
The more analytical approach, in the general case from the mathematical point of view, comes from the fundamental study of the epidemic growth and includes a number of terms in a form of double summation related to the inverse Laplace Transform of a rational function given in \cite{Kermack}, referring to the ``Earlier stages of an epidemic in a large population''. In this hypothesis, the number of unaffected individuals may be considered to be constant, while any alternation is assumed small compared to the total number of exposed individuals. This function, which could be called ``Large Population Epidemic Semi-Gaussian model'' (LPE-SG), is the following

\beq
c(t)=\sum_{i=1}^{N}\sum_{j=1}^{M}A_{ij}{{t}^{n_{ij}}}{{e}^{-t/\tau_{ij} }}
\eeq
\\ 
where $A_{ij}$ are arbitrary amplitudes, $n_{ij}$ are the degrees of the model (assumed fractional in a general case) and $\tau_{ij}$ are the time constants representing in our case a mean infection time respectively. Also, $M$ and $N$ are the finite number of terms to be included. It is easy to prove that the ``peaking time'' of the function of each term depends on the product of $n_{ij}$ by $\tau_{ij}$, that is, $t_{\mathrm{p_{ij}}}=n_{ij}\tau_{ij}$. 

In practice, the number of the required terms should be determined according to the shape of the data and the desired achievable accuracy. After investigation of the fitting performance we concluded that, at most, two terms of the above double sum are adequate for our purpose. Also, the cross terms, with indices $ij$ and $ji$, cannot help more the flexibility of the model. In particular, a) the degree of the model can ``cover'' any early or late smaller outbreak of the daily cases, while b) the mean infection time is a characteristic inherent parameter of the disease under study and thus should be essentially constant. For these reasons, the expression with one term was adequate in most of the cases, whereas, the 6 free parameters allow a very good flexibility for the fitting. Therefore, we can write

\beq
c(t)=\sum_{i=1}^{2}\sum_{j=1}^{2}A_{ij}t^{n_{ij}}e^{-t/\tau_{ij}}=A_1t^{n_1}e^{-t/\tau_1}+A_2t^{n_2}e^{-t/\tau_2}
\eeq 

For the fitting procedure, we have used two alternative tactics, based on either the daily model or on the cumulative integral of it. The decision depends on the goodness of the fit in each case based on the criterion of minimizing the $\chi^2/\mathrm{dof}$, as we have done in our previous work. \cite{Maltezos}. The starting date (at $t=0$) was one day before the first reported case (or cases). The cumulative parametrization model and the fitting model take the following forms respectively

\beq
C(t)=A_1{{\tau_1 }^{1+n_1}}\left[ \Gamma (n_1+1)-\Gamma_\mathrm{c}(n_1+1,t/\tau_1)\right]+ A_2{{\tau_2 }^{1+n_2}}\left[ \Gamma (n_2+1)-\Gamma_\mathrm{c}(n_2+1,t/\tau_2)\right]
\eeq

\beq
C(t)=p_1p_3^{1+p_2}\left[ \Gamma (p_2+1)-\Gamma_\mathrm{c}(p_2+1,t/p_3)\right]+p_4p_6^{1+p_5}\left[ \Gamma (p_5+1)-\Gamma_\mathrm{c}(p_5+1,t/p_6)\right] 
\label{cummulative_model}
\eeq
\\ 
where the symbol $\Gamma$ represents the gamma function and the $\Gamma_\mathrm{c}$ the upper incomplete gamma function at time $t$.

\subsection{Implementation strategy}

The above generalized mathematical model, has the advantage of providing more flexibility when the raw data include a composite structure or superposition of more than one growth curves which could be coexisting. This is possible to happen due to restriction measures  applied during the evolution of an epidemic. Regardless of the number of terms used, the obtained parameters must be well understood, in the sense of their role in the problem. Let us consider a country where the disease starts to outbreak due to a small number of infected individuals. In this stage we assume that the country has been isolated in a relatively high degree, but of course not ideally. At this point, the disease starts with a transmission rate which depends on the dynamics of the spread in each city and village, while other inherent properties of the disease affect its dynamics (e.g. the  immune reaction, the incubation time and recovering time). 

At this point, we must clarify also the issue of the ``size'' of the epidemic. The SIR-based models assume that the size, N, that is the total number of individuals exposable to the disease is unchanged during its evolution, a fact which cannot be exactly true. On the other hand, a fraction of the size concerns individuals who are in quarantine for different reasons (due to tracing or for precautionary reasons). Therefore, the size cannot be absolutely constant and the forecasting at the first stage (during the growing of the epidemic) should be very uncertain. In the second stage (around the turning point), the situation is more clear by means of more accurate parametrization, although high fluctuations could still be present. At the third stage (mitigation or suppress), an overall parametrization can be made and any trial for forecasting concerns a likely future comeback of the epidemic.
In any of the three stages regardless of the level of uncertainties the parametrization specifies the associated parameters according to the epidemic model used. It is known that the ``basic reproductive number'' symbolized by $R_0$ is a very important parameter of the spreading of the epidemic. In the SIR-based models it is determined at the first moments of the epidemic (mathematically at $t=0$) and is related to the associated parameters. Moreover, as it is proven in the next, this parameter doesn't depend on the size $N$ of the epidemic.  

By using the present parametrization model we assume that the size of the epidemic is, not only unknown, but also much smaller than the population of the country or city under study, that is, it constitutes an unbiased sample of the potentially exposable generic population. Once the epidemic is pretty much eliminated, the size could be also estimated ``a posteriori'' by the help of a SIR-based model. However, in this case, the parameters of the spread, as well as, the reproductive number are already determined by the methodology given in the next. We consider that this more generic approach facilitates the fitting process and improves the accuracy because of the existence of an analytical mathematically optimal solution. 

\section{Coupling with the SIR-based models}
\label{s3}
\subsection{Short description of SIR model}

The classical model for studying the spread of an epidemic, SIR, belongs to the Mathematical or State-Space category of models along with a large number of other types which are analytically described in \cite{Siettos_2}. Our model belongs to the continuum deterministic SIR models in the special case applied at the earlier stage of an epidemic, assuming that the population is much greater than the infected number of people \cite{Kermack}. This model can be applied also when the epidemic is in each latest stage where the total number of infected individuals is an unbiased sample of the population. Under these assumptions, this model can be related to the classical SIR model, or even with its extensions (SEIR and SIRD), by means of correlating their parameters. Below, we give a brief description of the basic SIR epidemic model.

Let us describe briefly the three state equations of the SIR model: 

\beq
\frac{\mathrm{d}S}{\mathrm{d}t}=-\frac{a}{N}SI
\label{S}
\eeq

\beq
\frac{\mathrm{d}I}{\mathrm{d}t}=\frac{a}{N}SI-\beta I 
\label{I}
\eeq

\beq
\frac{\mathrm{d}R}{\mathrm{d}t}=\beta I
\label{R}
\eeq

The function $S=S(t)$ represents the number of susceptible individuals, $I=I(t)$ the number of infected individuals and $R=R(t)$ the number of recovered individuals, all referred per unit time, usually measured in [days]. The constant factor $a$ is the transmission rate, the constant factor $\beta$ is the recovering rate and $N$ is the size of the system (the total number of individuals assumed constant in time), that is, $S+I+R=N$, at every time. The assumptions concerning the initial and final conditions are, $S(0)\ne 0,\mathrm{ }S(\infty )>0$, $I(\infty )=0$ and $R(0)=0,R(\infty )=N-S(\infty )$. 

This model does not have an analytical mathematical solution additional the two parameters $a$ and $\beta$ are constant during the spread of the epidemic. A solution is derived only with the approximation, $\beta R/N<1$, that is, when the epidemic essentially concerns a small number of recovered compared to the total number of individuals. In this case a Taylor's expansion to an exponential function is used. In our study, we work for the general case without this assumption.

\subsection{Synergy in the parametric space}

In our basic model, $C(t)$, the integral of the function $c(t)$, must be compared to the total number of infected individuals $I$, while the parameter $A$ undertakes the scaling of the particular data. The parameter $\tau$ does not necessarily coincides with the inverse of the `` mean infection rate'', $a$, but $1/\tau$ expresses an ``effective transmission rate'' in our model. The parameter $n$ cannot be equalized to any of the parameters of the SIR model. However, this parameter contributes to the key parameter of the epidemic spread, the so-called ``basic reproductive number'', $R_0$, which is defined at $t=0$ and is equal to $R_0=(a/\beta) (S(0)/N)$, where $\beta$ is the ``mean recovering rate''. Because $S(0)\approx N$, it becomes $R_0\approx a/\beta$. However, the $S(0)/N)$ represents a basic threshold, the so called ``population density'', above which the epidemic is initiated and growing when $R_o=(a/\beta)>1$.

Moreover the ``effective reproductive number'', $R_e$, a variable as a function of time is also defined by the same way as follows

\beq
{{R}_{e}}=-\frac{\mathrm{d}S}{\mathrm{d}R}=\frac{\mathrm{d(}I+R\mathrm{)}}{\mathrm{d}R}=1+\frac{\mathrm{d}I}{\mathrm{dt}}\frac{\mathrm{dt}}{\mathrm{dR}}=1+\left( \frac{a}{N}SI-\beta I \right)\frac{1}{\beta I}={{R}_{0}}\frac{S}{N}
\label{Re}
\eeq 

Because the condition for creating an epidemic is $R_e>1$, the corresponding condition should be $S/N>1/R_0$. Also, at $t=0$ should be $R_e(0)\equiv R_0$, at the peaking time $t=t_{\mathrm{p}}$ should be $R_e(t_{\mathrm{p}})=1$ and at $t=\infty$ takes the value ${{R}_{t}}={{R}_{0}}\frac{S}{N}(\infty)<R_0$. By using the expressions of Eq. \ref{Re}, including only the value of $I$ and its derivative, one can estimate roughly the $R_e$ at any time $t$. It can be also shown, that the $S(\infty )/N$ can be determined by solving the following transcendental equation numerically.

\beq
\frac{S(\infty )}{N}=\frac{S(0)}{N}{{e}^{-\frac{{{R}_{0}}}{N}\left[ N-S(\infty ) \right]}}\approx {{e}^{-{{R}_{0}}\left[ 1-\frac{S(\infty )}{N} \right]}}\approx {{e}^{-{{R}_{0}}\frac{R(\infty )}{N}}}
\label{s_infinity}
\eeq

Therefore, we can conclude that for $R_e$, at the outbreak of the epidemic (rising branch of the curve) we have, $1<R_e<R_0$, exactly at the peak of the curve, $R_e=1$ (because $S=N/R_0$, as we explain in the next) and at the mitigation stage (leading branch of the curve), $R_e<1$ and tends to a minimum value at the asymptotic tail of the curve which is 

\beq
{{R}_{e}}\approx {{R}_{0}}{{e}^{-{{R}_{0}}\frac{R(\infty )}{N}}}<<1
\eeq

Once the above relationship is achieved, the $R_0$ can be determined by solving the derived algebraic equation. Indeed, this was our initial motivation to perform the following analysis. The methodology for accomplishing it, was based on the idea to exploit the property of our model at its maximum at the peaking time, which is $t_\mathrm{p}=n\tau$, as it can be easily proven by differentiation. On the other hand, in the SIR model a peak is expected some time for the function $I$, as the typical effect of the epidemic's spread. Considering that both models can be fitted to the data, in the vicinity of the peak must agree, and therefore, we must claim that $I_{\mathrm{p}}=C(t_{\mathrm{p}})$. Let us first find an expression of the $S$, $R$ and $I$ at the peaking time, symbolizing them by, $S_{\mathrm{p}}$, $R_{\mathrm{p}}$ and $I_{\mathrm{p}}$ respectively. 

In order to relate $S$ with $R$, we  replace $I$ from Eq. \ref{S} into Eq. \ref{R}, we obtain

\beq
\ln S-\ln {{S}(0)}=-\frac{{{R}_{0}}}{N}\left[R-{{R}(0)} \right] 
\eeq
\\
from which, taking into account that $S(0)\approx N$, we derive the solution 

\beq
\ln \frac{S}{N}=-{{R}_{0}}\frac{R}{N}
\label{S_R_rel}
\eeq

The later result at the peaking time gives us an expression of $R_\mathrm{p}$

\beq
{{R}_{\mathrm{p}}}=-\frac{N}{{{R}_{0}}}\ln \frac{{{S}_{\mathrm{p}}}}{N}
\label{S_R_peak}
\eeq

Also, according to Eq. \ref{I}, at the peaking time we might have

\beq
\left( \frac{a{{S}_{\mathrm{p}}}}{N}-\beta  \right){{I}_{\mathrm{p}}}=0
\label{I_peak}
\eeq

From this equation, and using the definition of $R_0$, we obtain

\beq
{{S}_{\mathrm{p}}}=\frac{\beta N}{a}=\frac{N}{R_0}
\label{S_peak}
\eeq

Based on Eq. \ref{S_R_peak} we calculate $R_{\mathrm{p}}$ as follows

\beq
{{R}_{\mathrm{p}}}=-\frac{N}{{{R}_{0}}}\ln \frac{1}{{{R}_{0}}}
\eeq

Adding the three functions at the peaking time, $S_{\mathrm p}$, ${{I}_{\mathrm{p}}}$ and $R_{\mathrm p}$, we derive the algebraic equation 

\beq
\frac{{{I}_{\mathrm{p}}}}{N}+\ln {{R}_{0}}-{{R}_{0}}+1=0
\label{basic_eq}
\eeq

In order to achieve an equation independent of size $N$, we must express $I_{\mathrm{p}}/N$ as a function of the model parameters, that is, by using the maximum value of the model curve \cite{Maltezos} and the Eq. \ref{R} of the SIR model by integration with upper limit the infinity, as follows

\beq
\frac{{{I}_{\mathrm{p}}}}{N}=\frac{{{c}_{\max }}}{R\left( \infty  \right)+S\left( \infty  \right)}=\frac{{{c}_{\max }}}{R\left( \infty  \right)\left[ 1+\frac{S\left( \infty  \right)}{R\left( \infty  \right)} \right]}\approx \frac{A{{\left( \frac{n_0}{\tau_0 } \right)}^{n_0}}{{e}^{-n_0}}}{A\beta {{\tau_0 }^{1-n_0}}\Gamma (n_0+1)\left( 1+\frac{S\left( \infty  \right)}{N} \right)}=\frac{{{\tau_0 }'}}{\tau_0 }\frac{{{n_0}^{n_0}}{{e}^{-n_0}}}{\Gamma (n_0+1)\left( 1+\frac{S\left( \infty  \right)}{N} \right)}
\eeq
\\
where the symbol $\Gamma$ represents the gamma function, $n_0$ and $\tau_0$ are the particular values obtained by a fitting) and ${\tau_0 }'=1/\beta$. Replacing the above expression to Eq. \ref{basic_eq} and setting $s_N=S(\infty)/N$ we obtain

\beq
f({{R}_{0}};{{n}_{0}};{{\tau }_{0}};{{{\tau }'}_{0}})=\frac{{{{{\tau }'}}_{0}}}{{{\tau }_{0}}}\frac{{{n}^{{{n}_{0}}}}{{e}^{-{{n}_{0}}}}}{\Gamma ({{n}_{0}}+1)\left( 1+{{s}_{\mathrm{N}}} \right)}+\ln {{R}_{0}}-{{R}_{0}}+1=0
\label{sol_fun}
\eeq 

This transcendental equation can be solved only numerically for $R_0$ in which the combined unknown $s_\mathrm{N}$ is also found numerically by using again another transcendental Eq. \ref{s_infinity}, by using multiple iterations leading to a converging accurate solution within 12 loops. The parameter $n$ of the model is essentially the expresser of $R_0$, while the obtained value of $R_0$ concerns an hypothetical SIR model fitted to the data of the daily reported cases (DRC). From the obtained solution for $R_0$ we can also calculate the parameter $a$ of SIR model, $a=\beta R_0$, where $\beta$ can be calculated from the peak value of the daily reported recovered individuals by the formula, $\beta ={{R}_{\mathrm{p}}}/{{I}_{\mathrm{tot,p}}}$, where ${I}_{\mathrm{tot,p}}$ represents the integral of the DRC curve with upper limit the peaking time $t_{\mathrm{p}}$. In particular,

\beq
\beta =\frac{{{R}_{p}}}{A\tau _{0}^{^{1+n}}\left[ \Gamma ({{n}_{0}}+1)-{{\Gamma }_{\mathrm{c}}}({{n}_{0}}+1,{{n}_{0}}) \right]}
\eeq

\begin{figure}[!ht]
\centering
\begin{minipage}[b]{0.41\linewidth}
\centering
\includegraphics[scale=0.32]{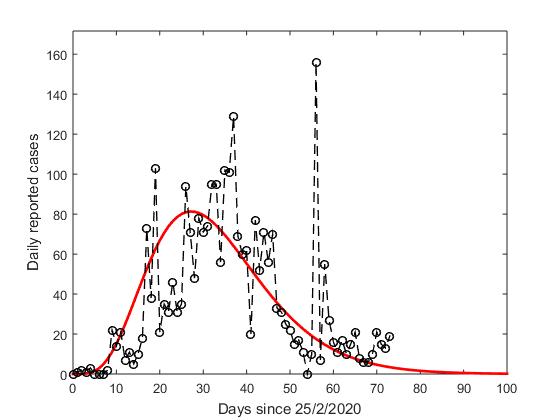}  
\caption[]{Indicative plot of the daily reported cases in Greece until May 9, 2020 (black circles) and the
optimal curve of the fitted model (red solid line) obtained by the total reported cases.}
\label{daily_GR}
\end{minipage}
\qquad
\begin{minipage}[b]{0.4\linewidth}
\centering
\includegraphics[scale=0.32]{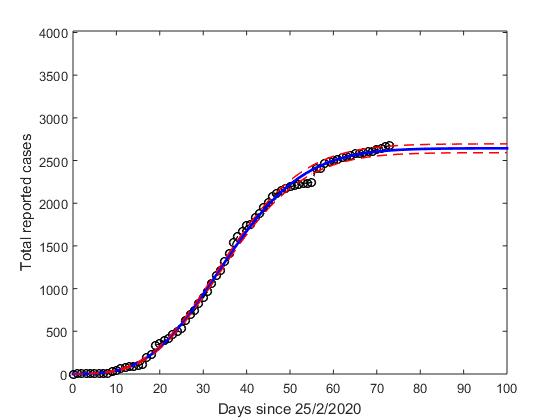}
\caption[]{The corresponding total reported cases in Greece (black circles) with the fitting (blue solid line). The uncertainty zone concerns $1\sigma$ and is shown by the red dashed lines.}
\label{total_GR}
\end{minipage}
\end{figure}

\begin{figure}[!ht]
\centering
\begin{minipage}[b]{0.42\linewidth}
\centering
\includegraphics[scale=0.32]{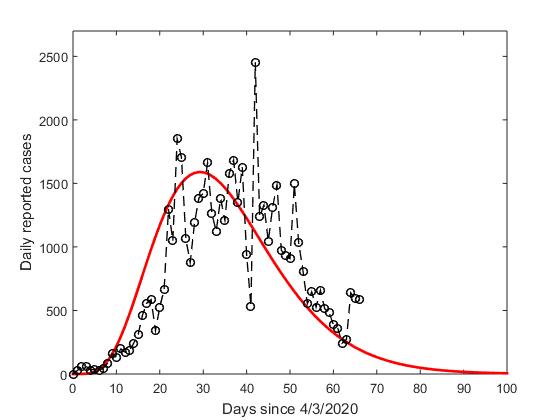}  
\caption[]{Indicative parametrization (red solid line) of the daily reported cases (black circles) in Belgium, as a typical example of a very large normalized size.}
\label{daily_BE}
\end{minipage}
\qquad
\begin{minipage}[b]{0.40\linewidth}
\centering
\includegraphics[scale=0.32]{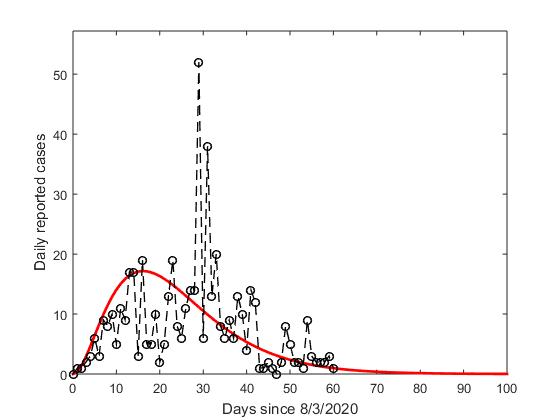}
\caption[]{Indicative parametrization (red solid line) of the daily reported cases (black circles) in Malta, as a typical example of a small normalized size.}
\label{daily_MA}
\end{minipage}
\end{figure}

Implementing the above methodology, by using home made software codes written in Matlab platform \cite{MATLAB}, we obtained the fitting of the DRC curve for Greece at the mitigation stage, shown in Fig. \ref{daily_GR} and Fig. \ref{total_GR}. The fitted parameters, $n_0=4.57$ and $\tau_0=5.96$ and the solutions $R_0=2.90$ and $s_\mathrm{N}=0.06$. For the parameter $\beta$ we used a typical-average value found in the literature where, $\beta=0.10$ and the same value we used for the other analyzed countries.
Two characteristic parametrizations of very large normalized size and very small one (64 times smaller), that is, of Belgium and Malta, respectively, are given in Fig. \ref{daily_BE} and Fig. \ref{daily_MA}. Definitely, without seeing the vertical scale, one cannot distinguish which corresponds a large or a small normalized size. The only visible difference at a glance, is the peaking time (29 and 16~days respectively).

\begin{table}[!h]
\centering
\begin{tabular}{|c|c|c|c|c|c|c|c|c|}
\hline 
COUNTRY &    $D$ $\mathrm{[p/km^2]}$ &  $\tilde{N}~[\mathrm{size/1Mp]}$ & GDP [Euros] & $n_0$ & $\tau_0 ~\mathrm{[days]}$ &  $R_0$  & $a=\beta R_0 ~\mathrm{[days^{-1}]}$ \\ 
\hline 
Austria          & 109   & 12000 & 44900 & 5.81 &  3.61  & 4.55  &  0.455 \\ 
\hline 
Belgium          & 383   & 115200 & 41200 & 4.77 &  6.13  &  2.78 & 0.277 \\ 
\hline
Bulgaria         & 64    & 1600 &  8680 & - &  -  &  - & -               \\ 
\hline
Croatia          & 73    & 2400 & 13330 & 2.99 &  6.14  &  3.47 &  0.347  \\ 
\hline
Cyprus           & 131   & 2400 & 24920 & 2.86 &  5.99  &  3.65 & 0.365   \\ 
\hline
Czechia          & 139   & 4000 & 20640 & 4.21 &  5.57  &  3.23 & 0.323   \\ 
\hline
Denmark          & 137   & 14000 & 53430 & 2.49 &  10.3  &  2.34 & 0.234  \\ 
\hline
Estonia          & 31    & 7000 & 21160 & 1.12 &  9.79  &  3.39 & 0.339   \\ 
\hline
Finland          & 18    & 6400 & 43480 & 2.18 &  12.5  &  2.14 & 0.214  \\ 
\hline
France           & 119   & 60200 & 36060 & 6.49 &  5.22  &  2.80 & 0.280  \\ 
\hline 
Germany          & 240   & 13800 & 41350 & 6.95 &  4.83  &  2.92 & 0.291  \\ 
\hline     
Greece           & 81    & 2400 & 17500 & 4.52 &  6.01  &  2.89 & 0.289  \\ 
\hline 
Hungary          & 107   & 5400 & 14720 & 4.38 &  7.41  &  2.45 & 0.245    \\ 
\hline
Ireland          & 72    & 41000 & 59910 & 4.38 &  5.09  &  3.49  & 0.349  \\ 
\hline
Italy            & 206   & 86000 & 29610 & 4.31 &  8.37  &  2.26 & 0.226   \\  
\hline
Lithuania        & 43    & 3000 & 17340 & 4.38 &  8.38  &  2.25 & 0.225  \\ 
\hline
Luxembourg       & 242   & 27200 & 102200 & 2.46 &  5.41  &  4.59 & 0.459  \\ 
\hline
Malta            & 1380  & 1800 & 26350 & 2.04 &  7.90  &  3.20 & 0.32   \\ 
\hline
Netherlands      & 508   & 50000 & 46820 & 4.54 &  7.28  &  2.46 & 0.246  \\ 
\hline
Poland           & 124   & 2600 & 13780 & 3.65 &  8.83  &  2.30 & 0.23   \\ 
\hline
Portugal         & 111   & 16800 & 20660 & 2.55 &  7.65  &  2.98 & 0.298  \\  
\hline 
Romania          & 84    & 6000 & 11500 & 2.85 &  10.0  &  2.28 & 0.228   \\  
\hline 
Serbia           & 100   & 3200 & 6590 & 6.90 &  4.67  &  3.03 & 0.303    \\  
\hline
Slovakia         & 114   & 600 &  17270 & 3.21 &  7.84  &  2.64 & 0.264  \\ 
\hline 
Slovenia         & 103   & 7600 & 22980 & 1.40 &  9.02  &  3.34 & 0.334  \\  
\hline 
Spain            & 94    & 96400 & 26440 & 2.88 &  9.50  &  2.36 & 0.236  \\  
\hline 
Sweden           & 25    & 42600 & 46180 & 2.93 &  12.9  &  1.95 & 0.195 \\  
\hline 
Switzerland      & 219   & 36800 & 69760 & 3.79 &  5.34  &  3.57 & 0.357  \\ 
\hline 
United Kingdom   & 281   & 57400 & 34190 & - &  -  &  - &  -   \\ 
\hline 

\end{tabular} 
\caption{Summary of the obtained parameters for 29 countries in Europe, after analysing all the available reported data. By $\tilde{N}$ we symbolize the epidemic size normalized to one million people. The epidemic curve in Bulgaria and UK had not yet reached clearly the turning point at the time of the present study.} 
\label{results}
\end{table}

\section{Study of the end-to-end epidemic growth}
\label{s4}
\subsection{Correlation searches}

For the data analysis we selected the 29 countries of EU, including Switzerland and UK obtained from \cite{worldometers}. The characteristics of the countries  relevant to our study are summarized in Table \ref{results}. In particular, we used the population density, the estimated normalized total number of infected individuals (determined by the number of deaths by using a typical constant factor) and the Gross Domestic Product (GDP), nominal per capita. The degree of correlation among the above characteristics and the modelling parameters, was studied by the ``theoretical pearson linear correlation coefficient'' given by the following formula

\beq
\rho(X,Y)=\frac{\mathrm{Cov}(X,Y)}{\sigma_x \sigma_y}
\label{corr_coef}
\eeq
\\
where $X$ and $Y$ are considered normal random variables, $\sigma_x$ and $\sigma_y$ are the corresponding standard deviations and $\mathrm{Cov}(X,Y)$ is their covariance. 
However, as it is done in practice, we calculated the ``sampling pearson coefficient'' (SPC), $r(X,Y)$, for $n$ observed random pairs $(X_i,Y_i,...,X_n,Y_n)$, where the $X$ can represent the first selected variable and $Y$ the second one.

\begin{table}[!h]
\centering
\begin{tabular}{|c|c|c|c|c|}
\hline 
CORRELATION PAIR &  SPC & P-VALUE & STATISTICALLY SIGNIFICANT, C.I. \\ 
\hline 
$D-\tilde{N}$             & 0.0645    & 0.749   & No   \\ 
\hline 
$\mathrm{GDP}-\tilde{N}$  & 0.311    & 0.113   &  No   \\ 
\hline 
$n_0-\tau_0$              & -0.645   & $<$0.001   & Yes, $99\%$ \\ 
\hline 
$D-R_0$                   & 0.101   & 0.617   & No   \\ 
\hline
$\tilde{N}-R_0$           & -0.193    & 0.335   & No \\ 
\hline
$t_{\mathrm{p}}-R_0$       & -0.734    & $<$0.001   & Yes, $99\%$  \\ 
\hline
$t_{\mathrm{p}}-\tilde{N}$ & 0.309    & 0.117   & No  \\ 
\hline
$t_{\mathrm{p}}-D$         & -0.169   & 0.399   & No  \\ 
\hline
\end{tabular} 
\caption{Summary of the correlation results.} 
\label{correlations}
\end{table}

The correlation study concerned eight pairs, as is illustrated in Table \ref{correlations}. The conclusions of the linear correlation study are the following:

\begin{enumerate}
\item{For the population density $D$: no correlation was found with other parameters.}
\item{For the model parameters $n_0$ and $\tau_0$: strong anti-correlation was found.}
\item{For the peaking time $t_{\mathrm{p}}$: very strong anti-correlation was found with the basic reproductive number, $R_0$.}
\end{enumerate}

\begin{figure}[!ht]
\centering
\begin{minipage}[b]{0.42\linewidth}
\centering
\includegraphics[scale=0.33]{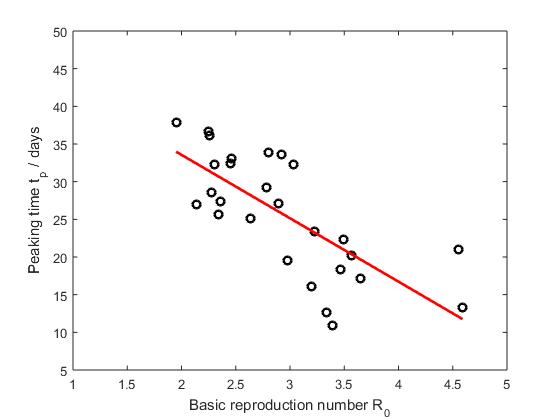}
\caption[]{A scatter plot of $R_0$ with the peaking time $t_{\mathrm{p}}$ and the linear fit.}
\label{scatter_tp_Ro}
\end{minipage}
\qquad
\begin{minipage}[b]{0.42\linewidth}
\centering
\includegraphics[scale=0.34]{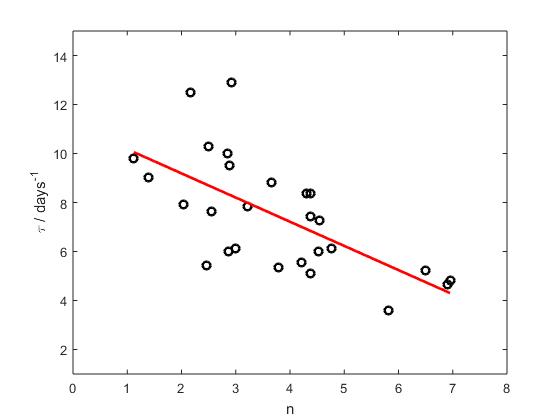}
\caption[]{A scatter plot of $n$ and $\tau$ and the linear fit.}
\label{scatter_n_tau}
\end{minipage}
\end{figure}

The scatter plot of the basic reproductive number $R_0$ and the peaking time $t_{\mathrm{p}}$ is shown in Fig. \ref{scatter_tp_Ro}. This correlation gives us the following message: the higher $R_0$ results to less a delay of the upcoming peak in the DRC curve. The obtained slope of the linear fit was $-8.4\pm0.2$~$\mathrm{days}$. On the other hand, the $R_0$ among the analyzed countries, present statistical fluctuations from about 2 to 4.6, obeying roughly a gaussian distribution with mean value $2.96$ and standard deviation $0.68$ (or relative to mean $23~\%$). The parameters $n$ and $\tau$ also fluctuate, as we can see in Fig. \ref{scatter_n_tau} while the peaking time $t_{\mathrm{p}}$ shows stochastic characteristics obeying similarly a gaussian distribution with mean $25.7$~days and standard deviation $7.8$~days (or relative to mean $30~\%$).

Since $R_0$ fluctuates (and in turn the $t_{\mathrm{p}}$ due to their linear correlation) among the different countries randomly without presenting any correlation with their associated parameters, we can conclude that the normalized size of the epidemic can be explained only by taking into account other reasons and aspects related to the way citizens interact and behave as well as the degree of social distances and mobility or transport within a country's major cities. Also, a crucial role played definitely the degree of quarantine and likely some individual biological differentiations (genetic and other related characteristics).

\subsection{Quantitative surveillance of the epidemic}
The capability for surveying the epidemic spread during the three main stages is very important and could be based on the daily data and the mathematical modelling we presented. In the mitigation stage the surveying is even more useful and crucial when the restriction measures are starting to be relaxed. The crucial condition for a new epidemic reappearance is based on the effective reproductive number, $R_e$, as well as, on the corresponding population threshold. However, because of the large statistical fluctuations caused by the poor statistics of data as well as because of the low slope of the epidemic curve in this stage, it is very hard to achieve accurate numbers, but only a qualitative estimate as follows. The $R_e$ can be estimated from the expressions in Eq. \ref{Re} and using average numerical approximations of the slope, $\mathrm{d}I/\mathrm{d}t$. An alternative and practical formula based on the parametrization model SG-LPE can be easily proven and is, $R_e=1+(n\tau/t-1)/\beta$. This formula is valid only in the vicinity of the peak, namely in the narrow range from $0.5t_p$ to $1.5t_p$, because the fitted model and the SIR one differ in the slopes at both side tails. Once $R_e$ is estimated, the population density threshold, in turn, can be calculated and should be $S(t)/N=1/R_e$, assuming that the normalized size $N$ can also be estimated by a similar level of accuracy. Therefore the crucial condition in the mitigation stage is written as

\beq
R_e>\frac{S(\infty)}{N}\Rightarrow 1+\frac{1}{\beta I}\frac{\mathrm{d}I}{\mathrm{d}t}>\frac{S(\infty)}{N}\Rightarrow \frac{\mathrm{d}I}{\mathrm{d}t}>\beta I\left( \frac{S(\infty)}{N}-1\right) 
\eeq     

The derivative has to be calculated as an average slope, $\mathrm{\Delta}I/\mathrm{\Delta}t$, preferably at least within one week. Assuming that this slope is $I'_{\mathrm{w}}$ and the corresponding average cases in a week is $I_{\mathrm{av}}$ the crucial condition becomes

\beq
I'_{\mathrm{w}}>\beta I_{\mathrm{av}}\left( \frac{N}{S(\infty)}-1\right)\approx 1.5I_{\mathrm{av}}
\eeq 
\\ 
where we used the typical values, $\beta=0.1~\mathrm{days}^{-1}$ and $\frac{N}{S(\infty)}\approx 15$ for having a practical result as a case study. This simplified formula combined with one-week measurements should be very useful because the relative fluctuations of the DRC are expected to be very large. 

\section*{Conclusions}
A systematic analysis of the epidemic characteristics of the new virus COVID-19 disease spread is presented in this work. For the mathematical analysis, we used a model that we called LPE-SG which facilitates the parametrization by an analytical mathematical description. We also presented a methodology of its coupling with the SIR-based models aiming to determine the basic and effective reproductive numbers based on the fitted parameters. Analysing the daily reported cases of European countries, we found no correlation between the population density, normalized size or GDP of the countries with respect to the spreading characteristics. Another important finding of our study was a strong anti-correlation, statistically significant, of the basic reproductive number and the peaking time. Moreover, we found that the basic reproductive number in the epidemics studied showed a uniform distribution with a wide range of values. This means that it is mainly influenced by many factors and generic characteristics of the society in a country.  

\noindent

\section*{Acknowledgements}

I would like to thank my daughter V. Maltezou, a Graduate of the Department of Agriculture of the Aristotle University of Thessaloniki and, of Athens School of Fine Arts, for our discussions on the global epidemiological problem, which gave me the warmth and the motivation for doing this work. Also, I thank my colleagues, Prof. Emeritus E. Fokitis and E. Katsoufis, for their insightful comments and our useful discussions.

\end{document}